\documentclass[reprint,showpacs,preprintnumbers,amsmath,amssymb,aps,prl]{revtex4-1}

\usepackage{graphicx}
\usepackage{dcolumn}
\usepackage{bm}
\DeclareMathOperator{\sinc}{sinc}

\begin{document}

\title[Wide Bandwidth, Frequency Modulated Free Electron Laser]{Wide Bandwidth, Frequency Modulated Free Electron Laser}
\author{L.T. Campbell$^{1,2}$}
\email[]{lawrence.campbell@strath.ac.uk}
\author{B.W.J. M$^{\mathrm c}$Neil$^1$}
\email{b.w.j.mcneil@strath.ac.uk}
\affiliation{1 SUPA, Department of Physics, University of Strathclyde, Glasgow, G4 0NG and\\ Cockcroft Institute, Warrington WA4 4AD, UK}
\affiliation{2 ASTeC, STFC Daresbury Laboratory, Warrington WA4 4AD, UK}

\date{\today}

\begin{abstract}
It is shown via theory and simulation that the resonant frequency of a Free Electron Laser may be modulated to obtain an FEL interaction with a frequency bandwidth which is at least an order of magnitude greater than normal FEL operation. The system is described in the linear regime by a summation over exponential gain modes, allowing the amplification of multiple light frequencies simultaneously. Simulation in 3D demonstrates the process for parameters of the UK's CLARA FEL test facility currently under construction. This new mode of FEL operation has close analogies to Frequency Modulation in a conventional cavity laser. This new, wide bandwidth mode of FEL operation  scales well for X-ray generation and offers users a new form of high-power FEL output.
\end{abstract}

\pacs{41.60.Cr}
\maketitle

The Free Electron Laser (FEL) is currently the world's brightest source of X-rays by many orders of magnitude~\cite{lcls,sacla,nprev}. The FEL consists of a  relativistic electron beam injected through a magnetic undulator with a co-propagating resonant radiation field. Initially,  co-propagating radiation will occur due to incoherent spontaneous noise emission from the electron beam and may be supplemented by an injected seed laser. The electrons can interact cooperatively with the radiation they emit and become density modulated at the resonant radiation wavelength. This coherently modulated oscillating electron beam exponentially amplifies the co-propagating radiation field in a positive feedback loop. In the single-pass high-gain mode, the energy of the initial incoherent, spontaneous X-rays may be amplified by around ten orders of magnitude. With such an increase in brightness over other laboratory sources, the X-ray FEL has unique applications across a wide range of the natural sciences.
FEL science is, however, still under development, and the creation of novel and improved output from the FEL is still an active topic of research. 

For example, it has been shown via simulations that equally spaced frequency modes may be generated in a single-pass FEL amplifier~\cite{mlsase,mlsaseab} by introducing a series of delays to the electron beam with respect to the co-propagating radiation field (e.g. by using magnetic chicanes placed between undulator modules).  These radiation modes are formally identical to those created in an oscillator cavity.  Analogously with a mode-locked conventional laser oscillator, a modulation of the electron beam energy~\cite{mlsase,mlsaseab} or current~\cite{popml} at the mode spacing can phase-lock the modes and amplify them to generate a train of short, high power pulses. 

Multiple colours may also be excited by directly tuning each undulator module to switch between 2 (or more) distinct colours \cite{sven2col}. This colour switching may also excite and amplify modes via a resulting gain modulation~\cite{margainmod}.

In what follows, the resonant frequency of the FEL is continuously modulated by varying the magnetic undulator field, as opposed to either a continual temporal modulation of \textit{e.g.} the beam energy, or a spatial variation between two distinct colours.

Control of the resonant frequency of the FEL $\omega_r$, may be achieved by varying the electron beam energy $\gamma_0$, the undulator wavenumber $k_u$, or undulator magnetic field strength $B_u$, via the resonance relation:
\begin{align}
\omega_r = \frac{2\gamma_0^2 ck_u}{(1 + \bar{a}_u^2)}, \label{resfreq}
\end{align}
where the mean undulator parameter $\bar{a}_u\propto B_u \lambda_u$, the product of its magnetic field and period respectively and $\gamma_0$ is the mean electron beam Lorentz factor. The undulator parameter $\bar{a}_u$ may be varied to change the resonant frequency as a function of the scaled propagation distance through the undulator $\bar{z}=z/l_g$ where $l_g=\lambda_u/4\pi\rho$ is the gain length of the FEL interaction and $\rho$ is the FEL parameter~\cite{FELexpsol}.

In this letter a periodic modulation of the FEL resonant frequency is made along the undulator length by varying $\bar{a}_u$. The spectral output is composed of equally spaced frequency modes, and it is found that the modulation amplitude extends the range of frequencies which may be amplified, allowing a control of the bandwidth within which the modes are amplified. 
This may be extended to orders of magnitude above the typical FEL amplification bandwidth, at the cost of lengthening the characteristic FEL gain length. The temporal radiation profile is seen to exhibit rapid temporal phase/frequency changes. 

\begin{figure*}
\centering
\includegraphics[width=17cm]{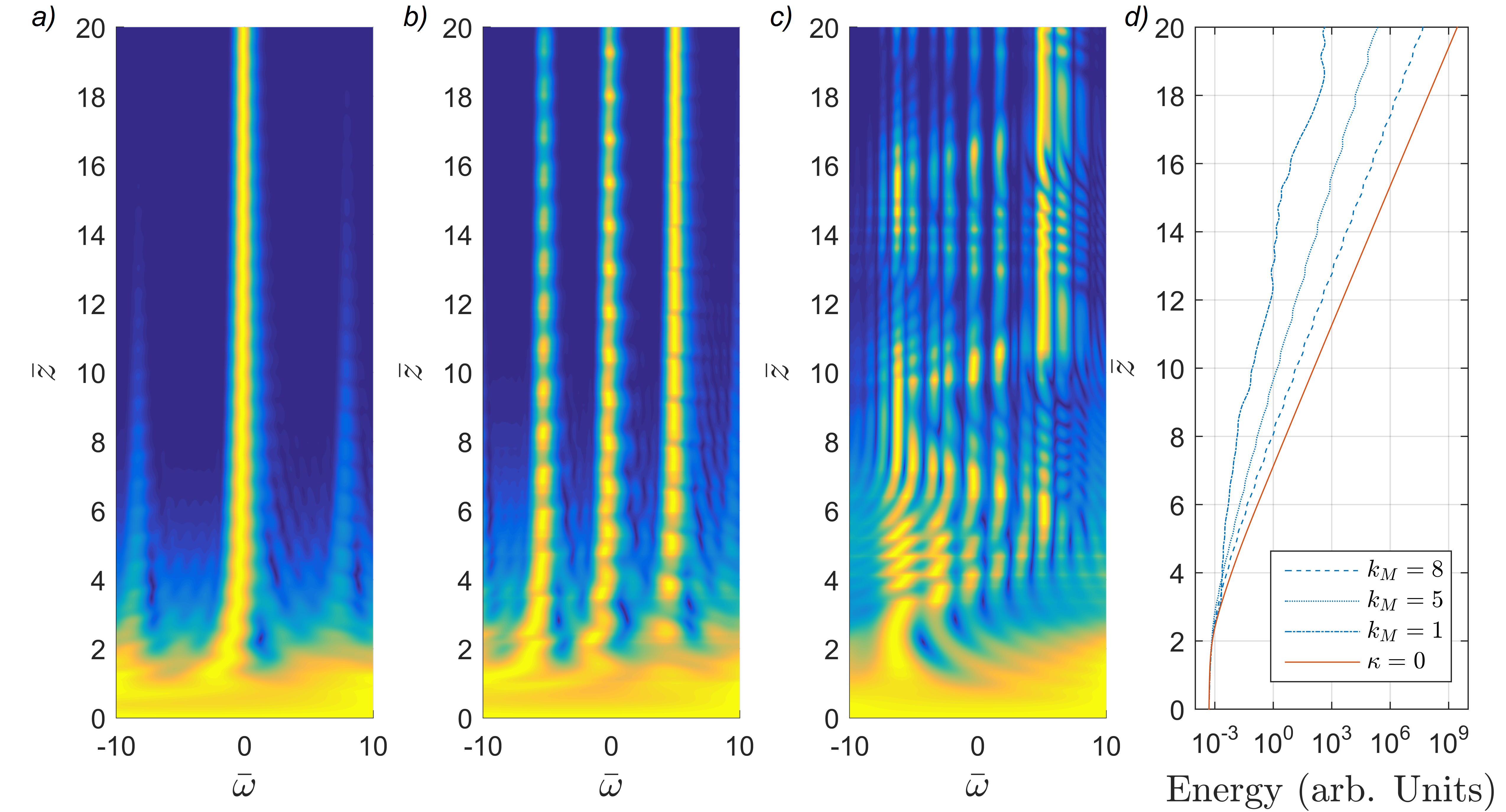}
\vspace*{-\baselineskip}
\caption[Sinusoidal Modulation]{Result of numerical solution of the linear system in eqns (\ref{bFT}-\ref{AFT}) using the sinusoidal frequency modulation, for $\rho = 0.001$ and $\kappa = 0.014$. The plots show the evolution of the normalised spectral intensity profile of the field from an initial broad bandwidth seed, constant for all frequencies. Each horizontal slice in $\bar{z}$ is normalised to its maximum: a) $\bar{k}_M = 8$; b) $\bar{k}_M = 5$; and c) $\bar{k}_M = 1$ corresponding to modulation index $\mu=0.875$, $1.4$ and $7$ respectively.}
\label{figure1}
\vspace*{-\baselineskip}
\end{figure*}

A system of linearised equations was used in~\cite{Bonifacio-taper} to investigate the efficiency of a tapered undulator FEL. Here a similar linear model is used but with a modulated undulator. Defining the undulator parameter to be  $\bar{a}_u(\bar{z}) = \bar{a}_{u0}(1 + \epsilon(\bar{z}))$, where the constant $\bar{a}_{u0} \gg 1$ and $|\epsilon(\bar{z})| \ll 1$, 
the following system of fourier-transformed, linearised FEL equations in $\tilde{b}(\bar{z}, \bar{\omega}), \tilde{P}(\bar{z}, \bar{\omega})$ and $\tilde{A}(\bar{z}, \bar{\omega})$ are obtained:
\begin{align}
\frac{\partial\tilde{b}}{\partial \bar{z}}  & = -i \tilde{P} \label{bFT} \\
\frac{\partial\tilde{P}}{\partial \bar{z}}  & = -\tilde{A} \label{PFT} \\
\frac{\partial\tilde{A}}{\partial \bar{z}}  & = \tilde{b} - i \left( \bar{\omega} + \epsilon\left(\bar{z}\right) \right) \tilde{A}, \label{AFT}
\end{align}
where $\bar{\omega}=(\omega/\omega_{r0}-1)/2\rho$ is the scaled radiation frequency, with $\omega_{r0}$ the radiation frequency for $\bar{a}_{u}=\bar{a}_{u0}$.
A modulation of the resonant frequency may be achieved within the above limits of validity via a sinusoidal modulation of the undulator parameter by setting $\epsilon(\bar{z}) = \kappa  \sin(\bar{k}_M \bar{z})/2$, where $\bar{k}_M$ defines the modulation period, to give $\omega_r(\bar{z}) / \omega_{r0}\approx 1 - \kappa \sin(\bar{k}_M \bar{z})$.

The spontaneous single electron output for such a modulated undulator, i.e.~with no FEL interaction,  may  be derived exactly and the solution for the scaled spectral power written as a sum over frequency modes with spacing $\bar{k}_M$: $|\tilde{A}(\bar{z}, \bar{\omega})|^2 = \bar{z}^2MM^*/2\pi$
where:
\begin{align} 
M=\sum_{n=-\infty}^{\infty} i^n e^{i\frac{\bar{\omega}_n \bar{z}}{2}} 
J_n\left( \frac{\omega}{\omega_{r0}} \mu \right)
\sinc\left(\frac{\bar{\omega}_n\bar{z}}{2}\right), 
\label{singleparticle}
\end{align}
with $\bar{\omega}_n\equiv \bar{\omega} + n \bar{k}_M$ describing a system of modes of scaled frequency spacing  $\bar{k}_M$. Here, $\mu=\kappa/2\rho\bar{k}_M$ is the modulation index, equivalent to that defined in conventional frequency-modulated lasers~\cite{FMsolidstate}. The modulation index gives an approximate upper limit to $|n|$ in equation~(\ref{singleparticle}) of those modes that fall within the modulated bandwidth, so that the number of modes present is $N_M\approx 2\mu+1$.

Performing a Laplace transform on~(\ref{bFT}-\ref{AFT}), ($\hat{A}(s,\bar{\omega})=\mathcal{L} \{\tilde{A}(\bar{z},\bar{\omega})\}$, etc) yields the following recursive solution of exponential modes in $s$:
\begin{align}
\hat{A}_n =\frac{s_n^2}{s_n^2 (s_n + \bar{\omega}) + 1} \left(\frac{\kappa}{4 \rho} \left(\hat{A}_{n-1} - \hat{A}_{n+1} \right) - \tilde{A}_i\right) \label{recursive}
\end{align}
where $s_n\equiv s + n \bar{k}_M$, $\hat{A}_n \equiv \hat{A}(s_n, \bar{\omega})$ and $\tilde{A}_i\equiv \tilde{A}(0,\bar{\omega})$. When $\kappa =n= 0$, this reduces to the  well-known cubic relation for the characteristic equation~\cite{FELexpsol}. 

The linear solution~(\ref{recursive}) indicates that the frequency modes of the radiation are not directly coupled - each  frequency evolves independently. Seeding just one of the radiation modes does not therefore couple to the other modes (although it should be noted that this model does not take into account coherent spontaneous emission~\cite{CSE}). However, the  \textit{spatial}, exponential modes (in $s$) are coupled, and result in a spatial gain modulation for each frequency mode. 

Taking the inverse Laplace transform of~(\ref{recursive}), the complex field envelope may be expressed as a convolution in $\bar{z}$ of the usual FEL gain function $G(\bar{z},\bar{\omega})$ with the frequency modulation:
\begin{align}
\tilde{A} =  - i G \ast \tilde{A} \frac{\kappa}{2\rho} \sin(\bar{k}_M \bar{z}) + G\tilde{A}_i,
\end{align}
where:
\begin{align}
G(\bar{z},\bar{\omega}) = \sum_{residues} \frac{s^2 e^{is \bar{z}}}{s^3 + \bar{\omega}s^2 + 1}
\end{align}
is the usual FEL gain function for a radiation seed. The `steady-state' FEL solution of~\cite{FELexpsol} is recovered for $\kappa = 0$.

\begin{figure}
\centering
\includegraphics[width=8.5cm]{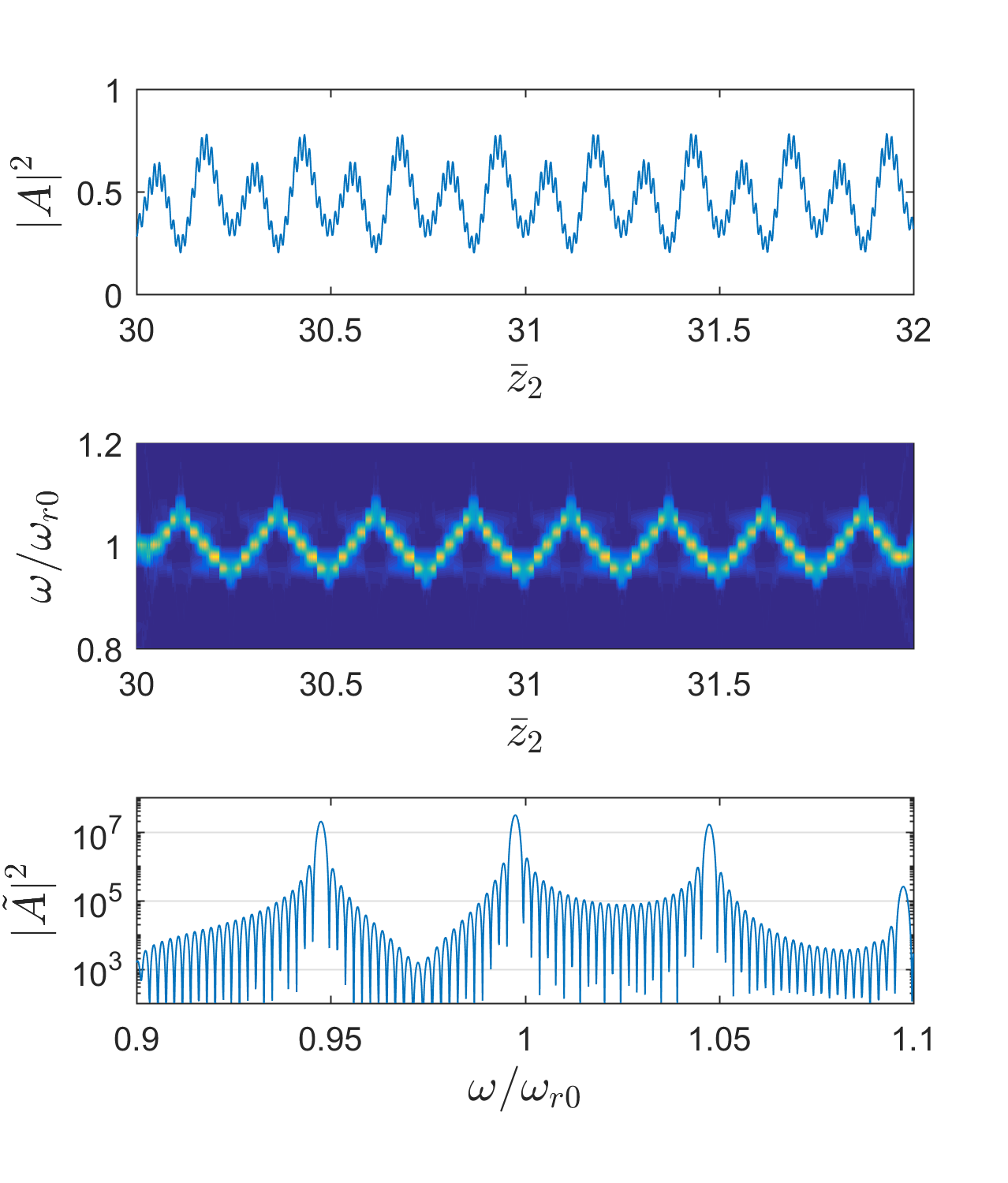}
\vspace*{-\baselineskip}
\caption[Sinusoidal Modulation]{Puffin simulation using a sinusoidal undulator modulation, for $\rho = 0.001$, $\kappa = 0.06$ and $\tilde{k}_M = 25$ at $\bar{z}=13.2$. Each mode is seeded with a scaled intensity $| A |^2 = 10^{-4}$. Top - scaled intensity vs $\bar{z}_2$, showing an almost CW profile, Middle - wavelet transform showing corresponding scaled frequency modulation, Bottom - fourier transform showing modal output. Note that radiation frequencies span 50 times the usual FEL spectral bandwidth.}
\label{figure2}
\vspace*{-\baselineskip}
\end{figure}

Results of numerical solutions of the linear system (\ref{bFT}-\ref{AFT}) are shown in Fig.~\ref{figure1} for three values of $\bar{k}_M$ and a range of the modulation index $\mu$ which covers few to several radiation modes. The system was seeded with a homogeneous broadband seed $\tilde{A}_i = 10^{-3}$. It is noted that the general form of the radiation emission of Fig.~\ref{figure1}, is  similar to that of the single particle emission of equation~(\ref{singleparticle}) which has a rich structure as a function of $\bar{z}$ and $\bar{\omega}$.
The undulator modulation generates radiation frequency modes spaced at the modulation frequency $\bar{k}_M$. For more than one set of sidebands within the modulated bandwidth the modulation index $\mu>1$. Furthermore, for the radiation modes to span a range greater than the usual FEL gain bandwidth, which is of interest in this letter, the modulation bandwidth $\kappa>2\rho$, or equivalently, $\mu\bar{k}_M>1$. Hence, as $\bar{k}_M$ decreases, more modes may fall within the extended bandwidth. 
Radiation pulse energy oscillations of period $2\pi/\bar{k}_M$ in $\bar{z}$  are seen to occur. For the $\bar{k}_M = 5$ case of Fig.~\ref{figure1}, each of the $3$ modes have a period in $\bar{z}$ of $\bar{\lambda}_M\approx 1.26$.

\begin{figure*}
\centering
\includegraphics[width=17cm]{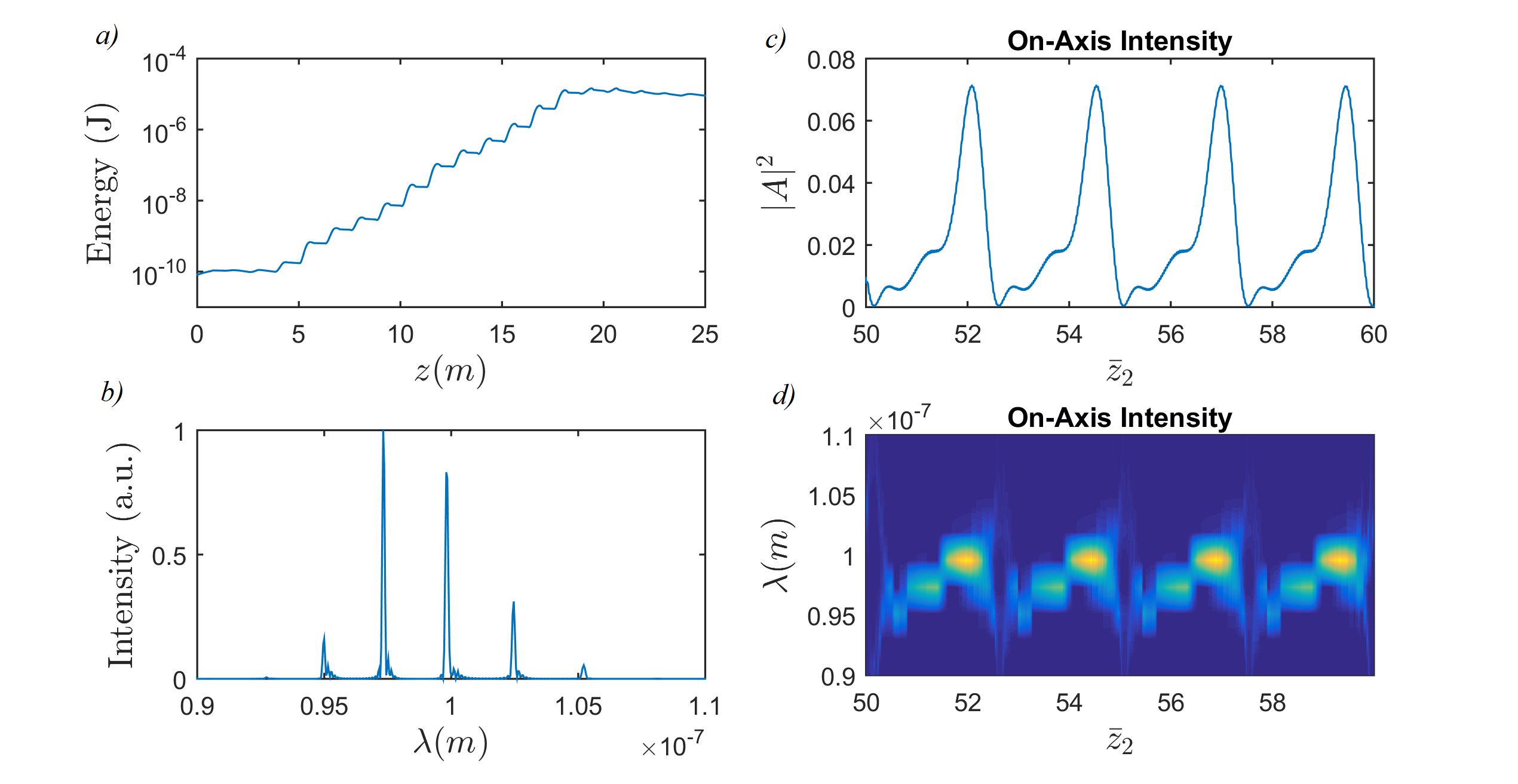}
\vspace*{-\baselineskip}
\caption[Sinusoidal Modulation]{ a) Radiated energy as a function of distance through the undulator line from CLARA simulation with modulated linear undulator tapers. The regions of no growth correspond to the spaces between undulator modules for the FODO focussing quadrupoles. Plots b), c), and d)  are respectively the spectrum, temporal power, and wavelength-temporal intensity correlation (by wavelet analysis) of on-axis radiation, just before saturation at $z \approx 17$m. The frequency modes have a spacing of $\approx 5 \rho$, within a bandwidth spanning $\approx 20 \rho$.} \label{figure4}
\vspace*{-\baselineskip}
\end{figure*}

Note that for the more extreme case of a modulation index of $\mu = 7$ shown in Fig.~\ref{figure1}~c),  exponential FEL gain occurs across many frequency modes. However, gain is not evenly distributed across the modes and being spread out over the many modes, occurs at a reduced rate from that of the usual unmodulated SASE case. Of course, this linear solution does not model saturation effects and in non-linear simulations using Puffin~\cite{puffin}, for the many mode cases examined, saturation was seen to occur at around twice the undulator length of the unmodulated SASE case.

In linear solutions of multi-mode cases one mode appears to dominate for $\bar{z} \gg 1$.
It also appears that frequency modes are driven even when non-resonant with the electrons as defined by the usual FEL relation~(\ref{resfreq}).
In non-linear models, saturation usually occurs before a single mode can completely dominate, leaving a modal structure with characteristic asymmetric Bessel-like `horns' familiar from conventional frequency modulated laser systems~\cite{FMsolidstate}.

Fig.~\ref{figure2} shows the radiation output from a 1D Puffin simulation in the non-linear regime with $\rho = 0.001$, $\kappa = 0.06$ and $\bar{k}_M = 25$ (modulation index $\mu=1.2$) close to saturation at $\bar{z} \approx 13.2$. 
An animation showing the evolution as a function of $\bar{z}$ through the undulator is available~\cite{video}. The temporal structure of the radiation power during the exponential growth of the linear regime (not shown) is composed of short  pulses ($\ll l_c$) due to the interference of the different modes. 
The modes do not have a fixed phase relationship. As the system saturates, the contrast of the radiation spikes reduces and their power structure briefly becomes more continuous with a strong frequency modulation, as seen in Fig.~\ref{figure2}. 
This is similar to that observed in a conventional FM laser \cite{FMsolidstate}. Post saturation, the pulsed output is seen to return, with a frequency modulation which further evolves in $\bar{z}$~\cite{video}. 
The large bandwidth of the amplified modes spans $\sim 100\rho$ but is achieved with a variation in $\bar{a}_u$ of only $3 \%$. In the X-Ray, where more typically the FEL parameter is reduced to $\rho \sim 10^{-4}$, a similar amplification bandwidth may be achieved with variations of $\bar{a}_u< 1 \%$.

For a practical applications, a periodic linear modulation to the undulator parameter $\bar{a}_u$ may be applied which has the same rate of frequency change for all frequencies. This version of the method is probably more easily realized at a short wavelength FEL facility, where the undulator is composed of many undulator modules, each of which would require the ability to be linearly tapered.
A simulation of this method using Puffin is shown in Fig.~\ref{figure4}. 
The parameters used are similar to the those of the future UK FEL test-facility CLARA~\cite{CLARA}, lasing at a wavelength $\lambda\approx 100$nm. 
The electron beam has energy $E = 350$MeV, normalised emittance $\epsilon_n = 0.5$mm mrad, and a flat-top current profile of duration $\sigma_t = 250$ fs and peak current $I_{pk} = 400$A. The planar undulators each have 25 periods of $\lambda_u = 2.75$cm and mean undulator parameter $\bar{a}_{u0} = 0.78$. 
The Puffin simulations are in 3D, include a FODO focusing lattice to maintain a mean beam radius of $\sim 60 \mu $m. The FEL interaction is seeded at each of the modes by an external radiation source of scaled intensity $|A|^2 = 10^{-4}$. 
The radiation has a relatively short Rayleigh range of $\approx 0.23$m, corresponding to approximately $1/3$ of an undulator module length, and therefore allowing a good test of the robustness of the method to 3D effects.
Each undulator module is linearly tapered from over a range $0.7\leq \bar{a}_u\leq  0.85$. As seen from the figure, 3D effects do not appear to adversely affect the frequency modulation in this simulation. Multiple modes are seen to be excited and amplified within a frequency bandwidth of approximately $20\rho$. The analysis of the linear system of equations showed that modes are driven even when strictly non-resonant. It would appear that there are diffractive guiding effects similar to that which occurs in normal, single frequency FEL operation. The wavelet analysis of Fig.~\ref{figure4}~d) and~\cite{video} shows a clear correlation of the radiation  frequency to the temporal power profile. 

The FM-FEL has been shown to exhibit similar properties to that of its conventional laser counterpart. It is a more complex system to study due to the temporal slippage of the radiation field through the co-propagating, amplifying electron beam. The general concepts and methodology used here may also be valid, at least within certain limits,  to similar cooperative many-body systems, such as the Collective Atomic Recoil Laser (CARL) instability~\cite{CARL}. FM-FEL operation appears to scale well in the X-Ray regime, where only relatively small changes of the undulator parameter within each undulator module are required for a notable extension of the mode-amplification bandwidth over that of a typical FEL. In order to demonstrate the rich behaviour of the FM-FEL, multi-coloured radiation seeds were used in the simulations to avoid noise in the output spectra which would mask the underlying physics. Note that such seeds could be provided by an HHG seed source in the soft X-Ray. In principle it may be possible to seed the FM-FEL interaction via e.g.~Echo-Enabled Harmonic Generation (EEHG)~\cite{EEHG}, or other future methods, in which the bunching spectra exhibits a plateaux of harmonics/modes. The use of chicanes~\cite{mlsase} between the tapered undulator modules may also allow further control of the the modes, such as independent control of both the amplification bandwidth of modes and the mode spacing. These methods could lead to a powerful, tunable, broadband source of multi-coloured coherent radiation into the hard X-ray.

\begin{acknowledgments}
We gratefully acknowledge
support of the Science and Technology Facilities
Council Agreement Number 4163192 Release \#3;
ARCHIE-WeSt HPC, EPSRC grant EP/K000586/1; EPSRC
Grant EP/M011607/1; and John von Neumann Institute
for Computing (NIC) on JUROPA at J\"ulich Supercomputing
Centre (JSC), under project HHH20.
\end{acknowledgments}

\end{document}